\documentclass[pre,aps,twocolumn,showpacs]{revtex4}
\usepackage{graphicx}
\usepackage{color}
\usepackage{epsf}
\usepackage{amsmath}
\usepackage{amssymb}
\usepackage[english]{babel}
\usepackage{bm}
\usepackage{times}
\usepackage[T1]{fontenc}
\usepackage{graphics}
\usepackage{color}
\usepackage{dcolumn}   
\usepackage{sidecap}
\usepackage{ulem}

\newcommand {\xx} {\ensuremath{{\bf x}}}

\newcommand {\ff} {\ensuremath{{\bf f}}}

\newcommand {\xxi} {\ensuremath{{\bf \xi}}}
\newcommand{\eps}{\epsilon}

\newcommand{\bes}{ \begin{equation} \begin{split} }
\newcommand{\ees}{ \end{split} \end{equation} }

\newcommand{\ignore}[1]{}

\renewcommand{\emph}[1]{{\it #1}}

\begin{document}

\title{Dynamical Bayesian Inference of Time-evolving Interactions: From a Pair of Coupled Oscillators to Networks of Oscillators}

\author{Andrea Duggento$^1$}%
\author{Tomislav Stankovski$^2$}%
\author{Peter V. E. McClintock$^2$}%
\author{Aneta Stefanovska$^2$}\email{aneta@lancaster.ac.uk}%

\affiliation{$^1$ Medical Physics Section, Faculty of Medicine, Tor Vergata University, Rome, Italy}
\affiliation{$^2$ Department of Physics, Lancaster University, Lancaster, LA1 4YB, United Kingdom}

\date{\today}

\begin{abstract}
Living systems have time-evolving interactions that, until recently, could not be identified accurately from recorded time series in the presence of noise. Stankovski et al.\ ({\it Phys.\ Rev.\ Lett.} {\bf 109} 024101; 2012) introduced a method based on dynamical Bayesian inference that facilitates the simultaneous detection of time-varying synchronization, directionality of influence, and coupling functions. It can distinguish unsynchronized dynamics from noise-induced phase slips. The method is based on phase dynamics, with Bayesian inference of the time-evolving parameters being achieved by shaping the prior densities to incorporate knowledge of previous samples. We now present the method in detail using numerically-generated data, data from an analog electronic circuit, and cardio-respiratory data. We also generalize the method to encompass networks of interacting oscillators and thus demonstrate its applicability to small-scale networks.
\end{abstract}


\maketitle

\section{Introduction}
\label{s:introduction}

Systems of interacting oscillators are ubiquitous in science. In the common case where the natural frequencies or amplitudes or inter-oscillator couplings are time-varying, they pose a continuing challenge to the time-series analyst who endeavours to understand the underlying system from the signal(s) it creates. Oversimplifications of hypotheses are often used to render the problem more tractable, but can all too easily result in a failure to describe phenomena that are in fact of central importance -- given that the strength, direction and functional relationships that define the nature of the interactions can cause qualitatively new states to appear or disappear. Time-variability of this kind is especially important in biological applications, though it is by no means restricted to biology.

In the absence of time-variability, there are many different methods available \cite{Palus:03a,Rosenblum:01,Bahraminasab:08,Jamsek:10} for detecting and quantifying the couplings and directionality (dominant direction of influence) between oscillators based, especially, on the analysis of phase dynamics. Approaches to the detection of synchronization have mostly been based on the statistical properties of the phase difference \cite{Tass:98,Mormann:00,Schelter:06,Xu:06}. The inference of an underlying phase model has been used as the functional basis for a number of techniques to infer the nature of the phase-resetting curves, interactions and structures of networks \cite{Galan:05,Kiss:05,Miyazaki:06,Kralemann:07,Tokuda:07,Levnajic:11}. However, these techniques inferred neither the noise dynamics nor the parameters characterizing the noise. An additional challenge to these methods can be the time-varying dynamics mentioned above. In a separate line of development, Bayesian inference was applied to analyse the system dynamics \cite{Smelyanskiy:05a,Friston:02,Sudderth:10,Luchinsky:08, Duggento:08,Penny:09}, thereby opening the door to inference of noisy time-evolving phase dynamics. Methods based on transfer entropy and Granger causality have a generality that has facilitated a number of applications, including inference of the coupling strength and directionality \cite{Daniele:08,Staniek:08,Hung:08}. These techniques provide measures of the amount of information in a measured signal, or the causal relationships between measured signals and, in doing so, they infer effect rather than mechanism.

In this paper we describe in detail and further extend a recently introduced method \cite{Stankovski:12b} based on dynamical Bayesian inference. As we demonstrate below, it enjoys many advantages over earlier approaches. One of these is that it does not require the observable to fill the domain of the probability density function at equilibrium: it can thus provide the same information from a small fraction of the data that is required by the transfer entropy or Granger causality approaches. Additionally, the dynamical approach has the advantage of characterizing the system completely (not only in terms of information measures). Thus, from the inferred dynamics one can deduce self-consistently any information that is of interest, be it coupling functions, or synchronization, or causality, or equilibrium densities, including the equations of motion. We discuss in detail the theoretical background, the technical aspects and limitations of the algorithms, and we demonstrate the wide applicability of the method by consideration of several examples.

The coupling functions are of particular importance. Their form is uniquely able to describe the functional laws of interaction between the oscillators. Earlier theoretical studies have included the work of Kuramoto \cite{Kuramoto:84} and Winfree \cite{Winfree:67}, which used a function defined either by the phase difference or by both phases, and of Daido \cite{Daido:96a, Daido:96b} and Crawford \cite{Crawford:95} who used a more general form in which the coupling function was expanded in Fourier series. Other methods for inference of the coupling functions have also been suggested \cite{Kiss:05,Miyazaki:06,Kralemann:08,Tokuda:07}. The technique described below goes beyond all of these because it is able to follow the time-variability of the coupling functions and hence can reveal their dynamical character where it exists.

We will also show how the technique can readily be extended to encompass networks of interacting oscillators. These form a large and important group of physical systems, including neural networks \cite{Galan:05,Cimenser:11,Deco:09}, electrochemical systems \cite{Kiss:05,Kiss:07}, crowd synchrony on the Millennium bridge, and networks of fireflies \cite{Strogatz:03b}. The large scale of the networks can introduce a higher complexity, both in structure and functional behavior. For example in neuronal networks, the existence of spatial and spatial-temporal correlations, collective or partially collective (clustering) behavior, synchronization or desynchronization, and time-variability has been reported \ignore{that}  \cite{Cimenser:11,Deco:09,Rudrauf:06,Tass:99}. In such cases, and given the kinds of phenomenon to be studied, there is an increasing need for powerful techniques that can infer the time-varying dynamics of the oscillatory networks.

In Sec.\ \ref{sc:dynamics} we provide details about the phase decompositions, the implementation of the Bayesian framework and how the time-varying information is propagated. The synchronization detection through a map representation of the phase dynamics is discussed in Sec.\ \ref{sc:synchdetection}, while the method for describing the interactions is demonstrated in Sec.\ \ref{sc:interactions}. Before the method is applied, we consider in Sec.\ \ref{sc:technical} some important technical aspects and limitations. The wide applicability of the method is demonstrated in Sec.\ \ref{sc:applications}, through the analysis of time-series from numerical phase and limit-cycle oscillators, analogue simulation and cardio-respiratory interactions. The generalization of the approach to networks of oscillators, as exemplified by two numerical examples, is presented in Sec. \ref{sc:networks}. Finally, we summarise and draw conclusions in Sec. \ref{sc:conclusion}. The algorithm used for the detection of synchronization is described in Appendix \ref{app:fixedpoint}.

\section{Phase-dynamics Decomposition} \label{sc:dynamics}

Consider an $N$-dimensional oscillator $d\xx/dt= \ff (\xx (t))$ whose solution $\ff$ admits a limit cycle. Such an oscillator can usually be represented by a constant phase velocity $\dot \phi=\omega$ and a vector coordinate that defines the limit cycle as a function of the phase $\phi$: ${\bf r} \equiv {\bf r} (\phi)$.

When two such oscillators mutually interact sufficiently weakly, their motion is commonly approximated just by their phase dynamics \cite{Kuramoto:84,Pikovsky:01}. We note that, in general, if we describe the phase of a system through a generic monotonic change of variables, than the dynamical process can be written as
\begin{equation}
\dot \phi_i= \omega_i + f_i(\phi_i) + g_i(\phi_i,\phi_j) + \xi_i.
\label{eq:phi}
\end{equation}
Eq.\ (\ref{eq:phi}) explicitly includes a noise term $\xi_i$ to
enable it to represent a process in a real system. The noise can
be e.g.\ a white Gaussian noise $\langle \xi_i(t)
\xi_j(\tau)\rangle = \delta(t-\tau) E_{ij}\ $, where the
symmetric matrix $E_{ij}$ encloses the information about
correlation between the noises on different oscillators, which we
will refer to as {\it spatial correlation}.

The phase-dynamics decomposition technique is highly modular from the algorithmic point of view, and each module will be explained separately in the sections that follow. The overall procedure comprises the following steps:

\begin{enumerate}

\item Assumption that the dynamics can be precisely described by a finite number of Fourier terms (see Sec. \ref{s:fourier}).

\item Inference, given the data, of the Fourier terms, the noise amplitude, and their correlation in form of a parameter probability distribution (see Sec. \ref{sec:bayesian} for stationary dynamics, and Sec. \ref{s:timevarying} for time-varying dynamics).

\item Integration of the probability that this parameter set lies inside the Arnold tongue defining synchronization. This effectively yields the cumulative probability of the synchronization state of the dynamics (see Sec. \ref{sec:discrimination}).

\item Use of the parameter information as obtained in step 2 to create a description of the interactions, leading to detection of the predominant directionality and coupling function estimation among the oscillators (see Sec. \ref{sc:interactions}).

\end{enumerate}

\subsection{Truncated Fourier Series}
\label{s:fourier}

The periodic behaviour of the system suggests that it can
appropriately be described by a Fourier decomposition. Decomposing
both $f_i$ and $g_i$ in this way leads to the infinite sums
\begin{eqnarray}
f_i(\phi_i) & = & \sum_{k=-\infty}^ \infty \tilde{c}_{i,2k} \sin(k\phi_i) + \tilde{c}_{i,2k+1}\cos(k\phi_{i}) \nonumber \\
{\rm and} \nonumber \\
g_i(\phi_i,\phi_j) & =  & \sum_{s=-\infty}^\infty  \sum_{r=-\infty}^\infty \tilde{c}_{i;r,s}\, e^{i2\pi r \phi_i}  e^{i2\pi s \phi_j}.
\label{eq:fourierdecomposition}
\end{eqnarray}
It is reasonable to assume that, in most cases, the dynamics will
be well-described by a finite number $K$ of Fourier terms, so that
we can rewrite the phase dynamics of Eq.(\ref{eq:phi}) as a finite
sum of base functions
\begin{equation}
\begin{split}
\dot \phi_i=& \sum_{k=-K}^{K} c^{(i)}_k \, \Phi_{i,k}(\phi_1,\phi_2)  + \xxi_i(t),
\label{eq:phiF}
\end{split}
\end{equation}
where $i=1,2$, $\Phi_{1,0}=\Phi_{2,0}=1$, $c^{(l)}_0=\omega_l$, and the rest of $\Phi_{l,k}$ and  $c^{(l)}_k$ are the $K$ most important Fourier components.

\subsection{Bayesian Inference}
\label{sec:bayesian}

In order to reconstruct the parameters of Eq.\ (\ref{eq:phiF}) we make extensive use of the approach to dynamical inference presented in \cite{Luchinsky:08,Duggento:08}. In this section we briefly outline the technique as adapted to the present case. The fundamental problem in dynamical inference can be defined as follows. A $2$-dimensional (in general $L$-dimensional) time-series of observational data ${\mathcal X} = \lbrace {\bf \phi}_{l,n} \equiv \phi_l(t_{n}) \rbrace$ ($t_n=nh$, $l=1,2$) is provided,
and the unknown  model parameters ${\mathcal M}=\{ c^{(l)}_k , E_{ij},\}$ are to be inferred.

Bayesian statistics employs a given \emph{prior} density $p_{\mbox{\scriptsize prior}}(\mathcal M)$ that encloses expert knowledge of the unknown parameters, together with a \emph{likelihood} function $\ell ( \mathcal X | \mathcal M )$, the probability density to observe $\{\phi_{l,n}(t)\}$ given the choice $\mathcal M$ of the dynamical model. Bayes' theorem
\begin{equation}
p_{{\mathcal X}}(\mathcal M | \mathcal X) = \frac{ \ell ( \mathcal
X | \mathcal M ) \, p_{\mbox{\scriptsize prior}}(\mathcal M) }{ \int{\ell (
\mathcal X | \mathcal M ) \, p_{\mbox{\scriptsize prior}}(\mathcal M) d
\mathcal M}  } \label{eq:bayes}
\end{equation}
then enables calculation of the so-called \emph{posterior} density $p_{{\mathcal X}}({\mathcal M}|{\mathcal X})$ of the unknown parameters ${\mathcal M}$ conditioned on the observations.

For independent white Gaussian noise sources, and in the mid-point approximation where
$$\dot \phi_{l,n}=\frac{{\phi}_{l,{n+1}}-\phi_{l,n}}{h} \, \, \,
 {\rm and}  \, \, \,  {\phi}_{l,n}^{\ast} = \frac{(\phi_{l,n} + \phi_{l,n+1})}{2}$$
the likelihood is given by a product over $n$ of the probability of observing $\phi_{l,{n+1}}$ at each time. The likelihood function is constructed by evaluation of the stochastic integral of the noise term over time, as
\begin{equation}
\xi^{(1)}_l(t_i) \equiv \int_{t_i}^{t_{i+1}} \xi_l(t) \, dt = \sqrt{h}\,H\, z_{l},  \,
\label{stochint}
\end{equation}
where $H$ is the Cholesky decomposition of the positive definite
matrix $E$, and $z_l$ is a vector of normally-distributed
random variables with zero mean and unit variance.
 The joint probability density of $z_l$ is used to find the joint probability
density of the process in respect of $(\phi_l(t_{i+1}) -
\phi_l(t_{i}))$ by imposing $P(\phi_l(t_{i+1})=
\det(J^{\phi}_{\xi})  P(\xi^{i})$, where  $J^{\phi}_{\xi}$ is the
Jacobian term of the transformation of variables that can be
calculated from Eq.\ (\ref{eq:fourierdecomposition}). If the
sampling frequency is high enough, the time step $h$ tends to
zero, and the determinant of the Jacobian $J^{\phi}_{\xi}$ can be
well-approximated by the product of its diagonal terms
$$
\det(J^{\phi_k(t_n)}_{\xi_k(t_n)}) \approx \prod_{l} \frac{\partial \Phi_{l,k}(\phi_{\cdot,n}) }{\partial \phi_{l}}\, .
$$
This transformation leads to an extra term in a least squares likelihood, and the minus log-likelihood function $S=-\ln \ell({\mathcal X}|{\mathcal M})$ can thus be written as
\begin{equation}
\begin{split}
    S &=   \frac{N}{2}\ln |{E}| + \frac{h}{2}\, \sum_{n=0}^{N-1}
     \Big ( c^{(l)}_k \frac{\partial \Phi_{l,k}(\phi_{\cdot,n}) }{\partial \phi_{l}}+\\
     &+ [\dot{\phi}_{i,n} - c^{(i)}_k {\Phi}_{i,k}({\phi}_{\cdot,n}^{\ast})] {({E}^{-1})}_{ij}  [\dot{\phi}_{j,n} - c^{(j)}_k {\Phi}_{j,k}({\phi}_{\cdot,n}^{\ast})] \Big ) \, ,
\end{split}
    \label{eq:likelihood}
\end{equation}
where summation over the repeated indices $k$,$l$,$i$,$j$ is implicit.

The log-likelihood (\ref{eq:likelihood}) is a quadratic form of the Fourier coefficients of the phases. Hence if a multivariate prior probability is assumed, the posterior probability is a multivariate normal distribution as well.

This is highly desirable for two reasons: (i) a Gaussian posterior is computationally convenient because
it guarantees a unique maximum, with the mean vector and covariance matrix completely characterizing the distribution and giving us the most significant information; (ii) all the multivariate normal posteriors can be used again as priors in the presence of a new block of data, and knowledge about the system can easily be updated. This last feature is essential for any real-time application because it ensures that the complexity of the algorithm does not change with the length of the input data-stream.

From \cite{Luchinsky:08}, and assuming a multivariate normal distribution as the prior for parameters ${c^{(l)}_{k}}$, with means $\bar {c}$, and covariances ${ \Sigma_{\mbox{\scriptsize prior}}}$, the stationary point of $S$ can be calculated recursively from
\begin{equation}
\begin{split}
    \label{eq:cD}
     E_{ij}  &= \frac{h}{N} \left(
 \dot{\phi}_{i,n} - c^{(i)}_k {\Phi_{i,k}}({\phi}_{\cdot,n}^{\ast})  \right)
\left(
 \dot{\phi}_{j,n} - c^{(j)}_k {\Phi_{j,k}}({\phi}_{\cdot,n}^{\ast}) \right) , \\
     c^{(i)}_k &= {({\Xi}^{-1})}^{(i,l)}_{kw} \,  {r}^{(l)}_{w} , \\
    {r}^{(l)}_{w}  & = {({\Xi}^{-1}_\text{prior})}^{(i,l)}_{kw} \,  {c}^{(l)}_{w} +
      h \, {\Phi_{i,k}}({\phi}_{\cdot,n}^{\ast}) \,
{(E^{-1})}_{ij} \, \dot{\phi}_{j,n} +\\
 &- \frac{h}{2} \frac{\partial \Phi_{l,k}(\phi_{\cdot,n}) }{\partial \phi_{l}}, \\
\Xi^{(i,j)}_{kw}  &= {\Xi_{\text{prior}}}^{(i,j)}_{kw}   + h \, {\Phi_{i,k}}({\phi}_{\cdot,n}^{\ast}) \,
{(E^{-1})}_{ij} \,   {\Phi_{j,w}}({\phi}_{\cdot,n}^{\ast}),
\end{split}
\end{equation}
where the covariance is ${\bf \Sigma={\bf\Xi}^{-1}}$, summation over $n$ from 1 to $N$ is assumed and summation over repeated indices $k$,$l$,$i$,$j$,$w$  is again implicit.

We note that a noninformative ``flat'' prior can be used as the limit of an infinitely
large normal distribution, by setting ${ {\bf\Xi}}_{\text{prior}}=0$ and $\bar c_{\mbox{\scriptsize prior}}=0$.

The multivariate probability ${\mathcal N}_{\mathcal X}(c|,\bar{c},\Xi)$ given the readout time series ${\mathcal X} = \lbrace {\bf \phi}_{l,n} \equiv \phi_l(t_{n}) \rbrace$ explicitly defines the probability density of each parameter set of the dynamical system. Because each of them can be discriminated, as belonging or not belonging to the Arnold tongue region we can define the binary property $ s(c^{(l)}_k)=\{1,0\}$, and can obtain the posterior probability of the system being synchronized or not by evaluating the probability of $s$
\begin{equation}
p_{\text{sync}} \equiv p_{\mathcal X}(s=1)= \int s(c)\, {\mathcal N}_{{\mathcal X}}(c|\bar c,\Xi) \, \text{d} c \, .
\label{eq:ps}
\end{equation}
The computation of $p_{\text{sync}}$ will be discussed in Sec. \ref{sc:synchdetection}.

\subsection{Time-varying information propagation}
\label{s:timevarying}

The multivariate probability described by ${\mathcal N}_{\mathcal X}(c,\Sigma)$ for the given time series ${\mathcal X} = \lbrace {\bf \phi}_{n} \equiv \phi(t_{n}) \rbrace$ explicitly defines the probability density of each parameter set of the dynamical system. When the sequential data come from a stream of measurements providing multiple blocks of information, one applies (\ref{eq:cD}) to each block. Within the Bayesian theorem, the evaluation of the current distribution relies on the evaluation of the previous block of data, i.e.\ the current prior depends on the previous posterior. Thus the inference defined in this way is not a simple windowing, but each stationary posterior depends on the history of the evaluations from previous blocks of data.

In classical Bayesian inference, if the system is known to be non-time-varying, then the posterior density of each block is taken as the prior of the next one: $\Sigma_{\text{prior}}^{n+1} =\Sigma_{\text{post}}^n$. This full propagation of the covariance matrix allows good separation of the noise, and the uncertainties in the parameters steadily decrease with time as more data are included.

\begin{figure}
\begin{center}
\includegraphics[width=0.98\linewidth,angle=0]{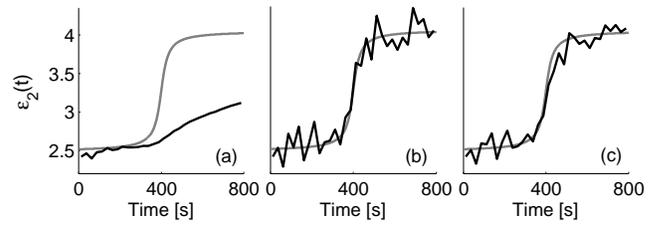}
\caption{Inference of a rapidly time-varying coupling parameter from coupled noisy oscillators (\ref{equ:num_model_app}). The gray lines represent the actual parameter in the numerical simulation, whereas the black lines indicate the time-varying parameter inferred from the resultant time series, for: (a) full propagation, $\Sigma_{\text{prior}}^{n+1} =\Sigma_{\text{post}}^n$; (b) no propagation, $\Sigma_{\text{prior}}^{n+1} =\infty$; and (c) propagation for time-varying processes, $\Sigma_{\text{prior}}^{n+1} = \Sigma_{\text{post}}^n + \Sigma_{\text{diff}}^n$ .  }\label{fig1:tv_propg}
\end{center}
\end{figure}

If time-variability exists, however, this propagation will act as a strong constraint on the inference, which will then fail to follow the variations of the parameters. This situation is illustrated in Fig.\ \ref{fig1:tv_propg}(a) \footnote{Note that Fig.\ \ref{fig1:tv_propg} shows inference of two coupled noisy Poincar\'{e} oscillators with the frequency of one oscillator time-varying  -- for clarity and compactness of presentation the details are not shown here, but the reader can refer to the model and other details in Sec.\ \ref{sc:applications}.}. In such cases, one can consider the processes between each block of data to be independent (i.e.\ Markovian). There cannot then be any information propagation between the blocks of data, and each inference starts from the flat distribution $\Sigma_{\text{prior}}^{n+1} =\infty$. The inference can thus follow more closely the time-variability of the parameters, but the effect of noise and the uncertainty of the inference will of course be much larger, as shown in Fig.\ \ref{fig1:tv_propg}(b).

Where the system's parameters are time-dependent, we may assume that their probability diffuses normally accordingly to the known diffusion matrix $\Sigma_{\text{diff}}$. Thus, the probability density of the parameters is the convolution of the two normal multivariate distributions, $\Sigma_{\text{post}}$ and $\Sigma_{\text{diff}}$
$$\Sigma_{\text{prior}}^{n+1} = \Sigma_{\text{post}}^n + \Sigma_{\text{diff}}^n.$$
The covariance matrix $\Sigma_{\text{diff}}$ expresses our belief about which part of the dynamical fields that define the oscillators has changed, and the extent of that change. Its elements are ($\Sigma_{\text{diff}})_{i,j}$ = $\rho_{ij}\sigma_i \sigma_j$, where $\sigma_i$ is the standard deviation of the diffusion of the parameter $c_i$ after the time window $t_w$ that has elapsed from the first block of information to the following one. $\rho_{ij}$ is the correlation between the change of the parameters $c_i$ and $c_j$ (with $\rho_{ii}=1$). In relation to the latter, a special example of $\Sigma_{\text{diff}}$ will be considered: we assume that there is no correlation between parameters, i.e.\ $\rho_{ij}=0$, and that each standard deviation $\sigma_i$ is a known fraction of the parameter $c_i$: $\sigma_i = p_w c_i$ (where $p_w$ indicates that $p$ is referred to a window of length $t_w$). It is important to note that this particular example is actually rather general because it assumes that all of the parameters (from the $\Sigma_{\text{post}}^{n}$ diagonal) can be of a time-varying nature -- which corresponds to the inference of real (experimental) systems with \emph{a priori} unknown time-variability.

There are two obvious limits in modeling the knowledge assumed with respect to
possible time variation of parameters. The first of these is to assume no
time-variability: in this case the full information propagation matrix is used,
$\sum_{\rm prior}^{n+1}=\sum_{\rm post}^n$. If the assumption proves wrong, the
inferred parameters may accumulate a bias when the real system varies in time.
The other limit is to assume each time window to be completely independent of
the previous signal history. In this case no propagation is used, $\sum_{\rm
prior}^{n+1} =\infty$, (i.e.\ $\Xi_{\rm prior}^{n+1} =0$), and there is no bias
but, because much information is forgotten, the probability of the inferred
parameters has a large covariance matrix. An optimal assumption must lies in
between these two limits: $\sum_{\rm prior}^{n+1} =\sum_{\rm post}^n+\sum_{\rm
diff}^n$ ; where the choice of $\sum_{\rm diff}^n$ is parameterized with the values of
the $p_w$'s. If a diffusion matrix is assumed, we allow the method some freedom
for the time-variability to be followed, while restricting it to be unbiased.
The amount of variability is part of the model, like the number of free
parameters in any standard method. Fig.\ \ref{fig1:tv_propg} illustrates the
two extreme limits, and a possible trade-off.
The inference in Fig.\ \ref{fig1:tv_propg}(c) demonstrates that the time-variability is captured correctly and that the uncertainty is reduced because more data have been included.

If one knows beforehand that only one parameter is varying (or, at most, a small number of parameters), then $\Sigma_{\text{diff}}$ can be customized to allow tracking of the time-variability specifically of that parameter. This selective propagation can be achieved if, for example, not all but only the selected correlation $\rho_{ii}$ from the diagonal has a non-zero value.

\section{Synchronization detection}\label{sc:synchdetection}

It is important to note that finite noise can induce phase slips in a system that would be synchronized in the noiseless limit.  Rather than focusing on the presence and statistics of phase-slips, we propose to detect synchronization from the nature of the phase-slip itself.  A novel feature of the present study is that it proposes evaluation of the probability that the equations driving the dynamics are {\it intrinsically} synchronized and thus of whether any phase-slips that may possibly be observed are dynamics-related or noise-induced.

After performing the inference, one can use the reconstructed
parameters, derived in the form of a multivariate normal
distribution ${\mathcal N}_{\mathcal X}(c,\Sigma)$, to study the
interactions between the oscillators under study. In general, the
border of the Arnold tongue may not have an analytic solution. In
practice, we estimate $p_{\text{sync}}$ numerically, sampling from
the parameter space many realizations $\{c^{(l)}_k\}_m$, where $m$
labels each parameter vector tested. For every set of $c$ we
compute $s(c_m)$ numerically. Let us assume for now that $s(c_m)$
is given. To find $p_{\text{sync}}$ with arbitrary precision, it
is enough to generate a number $M$ of parameters $c_m =
\{c^{(l)}_k\}_m$, with $m=1,\dots,M$ sampled from ${\mathcal
N}_{{\mathcal X}}(c|\bar c,\Xi)$, since $p_{\text{sync}}=
\lim_{M\to \infty}{\frac{1}{M}\sum_m^M} s(c_m)$.

However, this $2K$-dimensional integration quickly becomes inefficient with an increasing number of Fourier components. Moreover, as we will discuss in Sec.\ \ref{sec:discrimination}, the computation time of the variable $s(c_m)$ is not insignificant. On the other hand, if the posterior probability $p_{\mathcal X}$ is sharply peaked around the mean value $\bar c$, then $p_{\text{sync}}$ will be indistinguishable from $s(\bar c)$, and the evaluation of $s(\bar c)$ will suffice.

\subsection{Synchronization Discrimination and map representation}
 \label{sec:discrimination}

We now illustrate a simple numerical technique to recognize whether a coupled phase oscillator system is synchronized, or not. The technique itself amounts to a simple check by numerical integration of the system of ordinary differential equation defined by Eq.\ (\ref{eq:phi}) through one cycle of the dynamics, and testing whether the 1:1 synchronization condition $|\psi(t)|=|\phi_1(t)-\phi_2(t)|< K $ is always obeyed.

\label{sec:torus}

Let us assume we are observing motion on the torus $\mathbb T^2$
defined by the toroidal coordinate
$\zeta(\phi_1(t),\phi_2(t))=(\phi_1(t)+\phi_2(t))/2$, and the
polar coordinate $\psi(t)$.

\begin{figure}
\begin{center}
\includegraphics[width=4cm,angle=0]{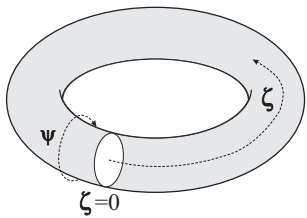}
\caption{Torus representation of the phase dynamics, with toroidal coordinate $\zeta(\phi_1(t),\phi_2(t))$ and polar coordinate $\psi(\phi_1(t),\phi_2(t))$. The white circle denotes the Poincar\'e cross section. }\label{fig2:torus}
\end{center}
\end{figure}

For assessment of possible 1:1 synchronization the phase
difference $\psi(t)$ will be defined as
$\psi(\phi_1(t),\phi_2(t))=\phi_1(t)-\phi_2(t)$. Fig.\
\ref{fig2:torus} provides a schematic representation of the phase
dynamics on the torus.
Let us consider a Poincar\'e section defined by $\zeta=0$ and
assume that $d\zeta(t)/dt |_{\zeta=0} > 0$ for any $\psi$. This
means that the direction of motion along the toroidal coordinate
is the same for every point of the section. Ideally we would
follow the time-evolution of every point and establish whether or
not there is a periodic orbit; if there is one, and if its winding
number is zero, then the system is synchronized. If such a
periodic orbit exists, then there is at least one other periodic
orbit, with one of them being stable and the other unstable.

\begin{figure*}
\begin{center}
\includegraphics[width=13.5cm,angle=0]{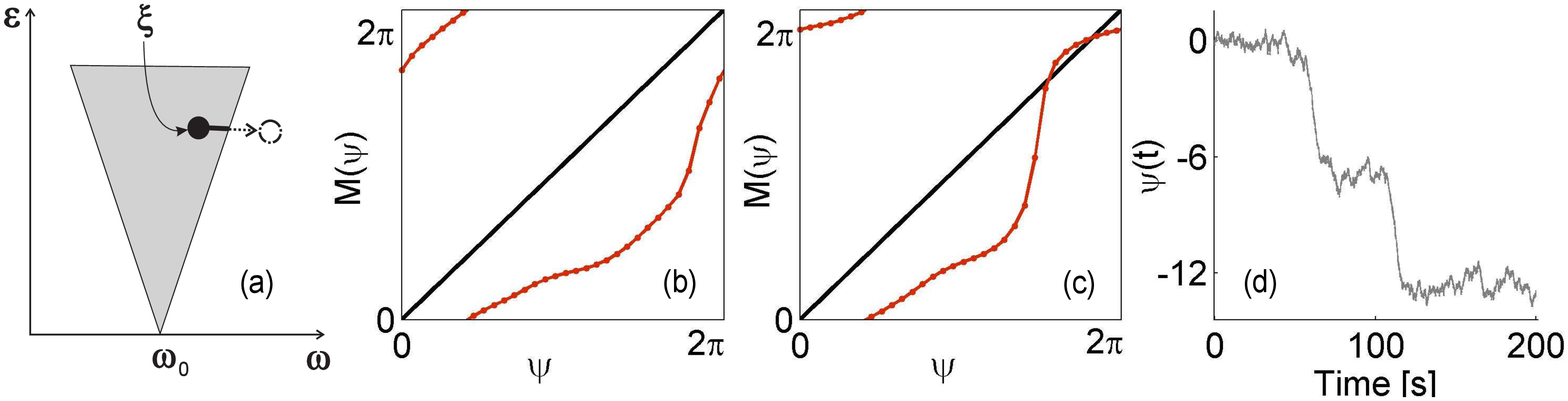}
\caption{(Color online) Synchronization discrimination for the coupled phase oscillators (\ref{eq:ph_app}). (a) Schematic of an Arnold tongue in the coupling-frequency $\varepsilon$-$\omega$ plane: synchronization exists only within the shaded area \cite{Pikovsky:01}. (b) Map of $M(\psi)$ for $\eps_{12}=0.25$ demonstrating that the oscillators are not synchronized. (c) Map of $M(\psi)$ for a case where a root of $M(\psi)=\psi$ exists, i.e.\ where that the state is synchronized. (d) The corresponding phase difference, exhibiting two phase slips. }\label{fig8:maps}
\end{center}
\end{figure*}

The solution of the dynamical system over the torus yields a map
$M:[0,2\pi]\to [0,2\pi]$ that defines, for each $\psi_n$ on the
Poincar\'e section, the next phase $\psi_{n+1}$ after one circuit
of the toroidal coordinate $\psi_{n+1}=M(\psi_n)$. Fig.\
\ref{fig8:maps}(b),(c) illustrates the map $M$ as evaluated
computationally in two situations, corresponding to no
synchronization, or synchronization, respectively.

The map $M$ is continuous, periodic, and has two fixed points (one stable and one unstable) if and only if there is a pair of periodic orbits for the dynamical system, i.e.\ synchronization is verified if $\psi_e$ exists such that $\psi_e=M(\psi_e)$ and $ \left| {\frac{dM(\psi)}{d \psi}}|_{\psi_e} \right| < 1$. The existence of the fixed point $\psi_e$ is established through the simple algorithmic procedure described in Appendix \ref{app:fixedpoint}.

\section{Description of the interactions}\label{sc:interactions}

Inferring the parameters of the system not only allows for evaluation of the synchronization as an epiphenomenon in its own right, but their probability ${\mathcal N}_{\mathcal X}(c,\Sigma)$ also describes the interaction properties of the oscillators. Because their dynamics is reconstructed separately, as described by Eq.\ (\ref{eq:phi}), use can be made only of those inferred parameters from the base functions $f_i(\phi_j)$ and $g_i(\phi_i,\phi_j)$ that are linked to influences between the oscillators.

One can seek to determine the properties that characterize the
interaction in terms of a strength of coupling, predominant
direction of coupling, or even by inference of a coupling
function. The analysis of information propagation allows inference
of the time-varying dynamics, and the interactions' properties can
be traced in time as well. This is especially important for the
inference of open interacting oscillatory processes where the
time-variability of the interactions can lead to transitions
between qualitatively different states, such as synchronization
\cite{Pikovsky:01}.

The coupling amplitude quantifies the total influence between the
oscillators in a particular direction, e.g.\ how much the dynamics
of the first oscillator affects the dynamical behavior of the
second oscillator ($1 \rightarrow 2$). Depending on whether the
coupling is in only one direction, or in both directions, we speak
of unidirectional or bidirectional coupling, respectively.  In the
inferential framework that we propose, the coupling amplitudes are
evaluated as normalized measures, based on the interacting
parameters inferred from the coupling base functions.  The
influence of one oscillator on the other can either be direct
through $f_i(\phi_j)$, or can arise through the combined
interacting base functions $g_i(\phi_i,\phi_j)$. In what follows,
the  base functions $f_i(\phi_j)$ and $g_i(\phi_i,\phi_j)$ are
described with a common notation $q_i(\phi_i,\phi_j)$. The
quantification is calculated as a Euclidian norm:
\begin{equation}
\begin{split} \eps_{21} &= \|q_1(\phi_1,\phi_2) \|\equiv\sqrt{c_1^2+c_3^2+\ldots} \\ \eps_{12} &= \|q_2(\phi_1,\phi_2) \|\equiv\sqrt{c_2^2+c_4^2+\ldots},\label{eq:norm}
\end{split}
\end{equation}
where the odd inferred parameters are assigned to the base
functions $q_1(\phi_1,\phi_2)$ for the coupling that the second
oscillator imposes on the first ($\eps_{21}: 2 \rightarrow 1$),
and {\it vice versa} ($\eps_{12}: 1 \rightarrow 2$).

The directionality of coupling \cite{Rosenblum:01} often provides
useful information about the interactions. It is defined as
normalization about the predominant coupling amplitude
\begin{equation}
D=\frac{\eps_{12}-\eps_{21}}{\eps_{12}+\eps_{21}}. \label{eq:dirc}
\end{equation}
If $D\in(0,1]$ the first oscillator drives the second ($1
\rightarrow 2$), or if $D\in[-1,0)$ the second ($2 \rightarrow 1$)
drives the first. The quantified values of the coupling strengths
$\eps_i$ or the directionality $D$ represent measures of the
combined relationships between the oscillators. Thus, a non-zero
value can be inferred even when there is no interaction. Such
discrepancies can be overcome by careful surrogate testing
\cite{Schreiber:96b,Palus:08a} -- by rejection of values below an
surrogate acceptance threshold, which can be specified e.g.\ as
the mean plus two standard deviations among many realization of
the measure.

In addition to the coupling strength and the directionality, one can also infer the coupling function that characterizes the interactions, i.e.\ the law that describes the functional relationships between the oscillators. Its characteristic form reflects the nature of the oscillators and how their dynamics reacts to perturbations.

The coupling function should be $2\pi$-periodic. In the
inferential framework under study, the coupling functions were
decomposed into a finite number of Fourier components. The
function describing the interactions between the two oscillators
was decomposed by use of the odd parameters
$q_1(\phi_1,\phi_2)\in\{c_1,c_3,\ldots \}$ and the corresponding
base functions $\Phi_n[q_1(\phi_1,\phi_2)]\in$
$\{\sin(\phi_1,\phi_2),$ $\cos(\phi_1,\phi_2)\}$ up to order $n$
of the decomposition. The reverse function $q_2(\phi_1,\phi_2)$
$\in$ $\{c_2,c_4,\ldots \}$ was similarly decomposed.

\section{Technical aspects and considerations}\label{sc:technical}
The technique is quite generally applicable to a broad class of problems, and so there are a number of technical aspects and choices to bear in mind. We now discuss three of them in particular: the number of base functions to be employed in the inference process (which is part of the model); the intensity of the noise characterizing the system (which is an externally imposed constraint); and the time resolution.

\paragraph{Number of base functions.} Selection of the optimal set of base functions to describe the problem is far from trivial. In general one wishes to have the minimal set that describes sufficiently well the model to be tested. Where the length of the data series is very long or effectively infinite, one can include an excessive number of base functions without immediate penalties. In reality, however, any unneeded  base function jeopardizes the precision of the coefficients that really are relevant for the model, and the picture is further complicated when the model to be adopted is expected to be an outcome of the inference machinery. Where one deals with a long data series, possibly with a high signal-to-noise ratio, a relative large number of base functions can be used. The speed of computation is also an important aspect to keep under consideration, given that having a large number of base functions vastly increases the parameter space, and that iterative calculations (especially matrix inversion) slow the speed of processing by the third power of the number of coefficients.  Note that, even though the Bayesian inference is generally popular in real-time applications, computational speed limitations mean that our inference framework for general phase dynamics cannot yet be used in this way.

\begin{figure}[b]
\begin{center}
\includegraphics[width=0.68\linewidth,angle=0]{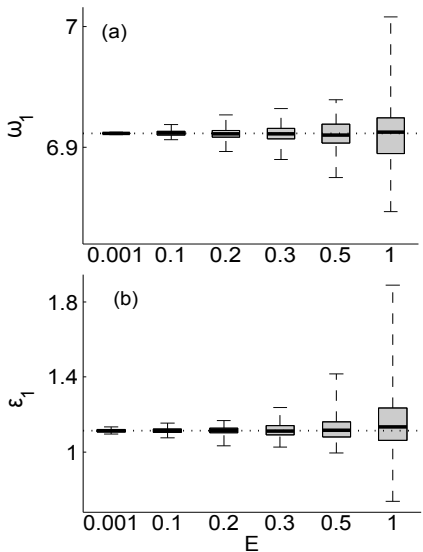}
\caption{Statistics of the inferred frequency $\omega_1$ and coupling $\varepsilon_{1}$ for different noise intensities $E$. The signal to be analysed is generated from Eq.\ (\ref{equ:num_model_app}).  The dotted line shows the actual values of the parameters.  The boxplots refer to the descriptive statistics (median, quartiles, max. and min.) of $10^4$ different runs of the generation-and-inference loop.}
\label{fig5:noises}
\end{center}
\end{figure}

\paragraph{Role of noise intensity.}\label{sec4_sub2_cph3:base_noises} In general, the greater the noise intensity, the bigger the covariance of the inferred parameters. For a repeated experiment (e.g.\ generation of a synthetic signal, and parameter inference based on that signal) the variance of a particular parameter would increase monotonically with noise amplitude, as shown in Fig.\ \ref{fig5:noises}.
There are, however, a few notable exceptions. The inferential capabilities rely on the volume of phase-space spanned by the variables.  A state of synchronization would represent a limit cycle for the global system, and parameter inference of neither oscillator  would reach satisfactory precision.  In such cases a minimal amount of noise is typically needed, sufficient to drive the system out of equilibrium at least once. During the resultant phase-slip, the data would be filling the phase space sufficiently for correct parameter reconstruction.

\paragraph{Time resolution.}\label{sec4_sub3_cph3:base_time_res} We now summarize the limits of an idealized data acquisition. The time step $h$ is much smaller than any of the sequential time-windows used as data blocks for inference, so that each block contains many data points. Also, $h$ is much smaller than either of the oscillator periods.  Each inference block is big enough to contain many cycles of the dynamics (in particular, more cycles that those typical of a phase slip) while, at the same time, each block is small enough to provide the desired resolution of
parameter change.

It can happen that the time resolution of the change in dynamical parameters is incompatible with an acquisition time-window that would guarantee precision for other parameters. The choice of the time-window must therefore be done on a case-by-case basis, depending on the type of information that is required from the system.
\begin{figure}[t!]
\begin{center}
\includegraphics[width=0.84\linewidth,angle=0]{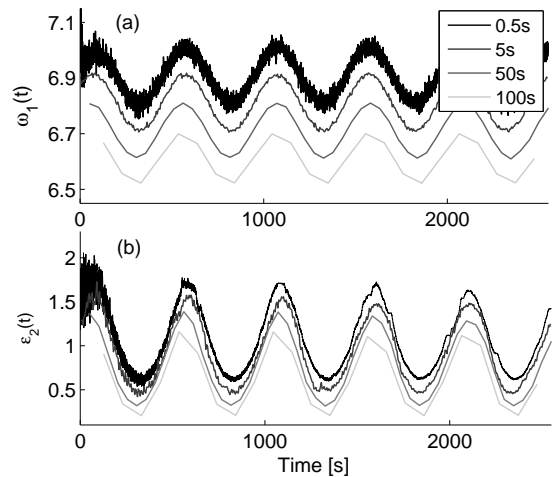}
\caption{Inference of (a) a time-varying frequency and (b)
coupling parameter from time series data generated by model
(\ref{equ:num_model_app}) for four different lengths of the
inference windows. The sizes of the windows are shown in the box.
For clearer presentation, the curves are separated by
shifting them vertically from each other through equal offsets in
$y$.}\label{fig6:time_res}
\end{center}
\end{figure}
Fig.\ \ref{fig6:time_res} illustrates such a compromise.  We use
the numerical model (\ref{equ:num_model_app}) that will be introduced in
Sec.~\ref{sec5_sub2_cph3:app_num} to investigate the time-resolution for the case where the frequency $\omega_1(t)=\omega_1+\tilde A_1
\sin (\tilde \omega t)$ and coupling amplitude $\varepsilon_{2}(t)=\varepsilon_{2}+\tilde A_1 \sin (\tilde \omega t)$ were varying periodically at the same time. The parameters were: $\omega_1=2\pi\,1.1$, $\omega_2=2\pi\,2.77$, $\varepsilon_{1}=0$, $\varepsilon_{2}=1$, $\tilde \omega=2\pi\,0.002$, $\tilde A_1 = 0.1$ $\tilde A_2 = 0.5$ and noise strengths $E_1=E_2=0.15 \,$. The parameters were reconstructed using four different window lengths for the inference. The results presented in Fig.\ \ref{fig6:time_res} demonstrate that, for small windows (0.5s), the parameters are sparse and sporadic, while for very large windows (100s) the time-variability is faster than the size of the window and there is cut-off on
the form of the variability. The optimal window length will lie between these two. Another interesting feature is that, for the smallest window (0.5\,s), the coupling amplitude improves with information propagation as time progresses, while the frequency inferred (as a constant component without base function) remains sparse throughout the whole time interval.

\section{Applications}
\label{sc:applications}
The technique is first applied to synthetic data to test the performance of the algorithm, and then real data are analyzed. To create the synthetic data, we used both numerical and analogue electronic simulations.  In the examples that follow, except where stated otherwise, we used the phase model Eqs.\ (1)-(3) with Fourier expansion to second order $K=2$, propagation constant $p_w=0.2$ and window length of $t_w=50$\,s.

\subsection{Numerically-generated test data}\label{sec6:numericaldata}

Numerically generated data were obtained from models of phase oscillators and limit-cycle oscillators.

\subsubsection{Phase oscillators}\label{sec:app_phs}
The phase oscillator model provides a sufficient basis for the description of synchronization while being, at the same time, analytically traceable. We thus test the detection of synchronization (as explained in Sec.\ \ref{sc:synchdetection}) through Bayesian inference of synthetic data whose synchronization is already known. The model is given by two coupled phase oscillators subject to white noise
\begin{equation}
\dot{\phi_i}=\omega_i  +\eps_{ji} \sin(\phi_j-\phi_i) + \xi_i(t) \,\,\,\,\;\;\; i,j=1,2 \, .
\label{eq:ph_app}
\end{equation}
The parameters are $\omega_1=1.2$, $\omega_2=0.8$, $\eps_{21}=0.1$; parameter $\eps_{12}$ is chosen so that the system lies close to the border of the Arnold tongue (either just inside or just outside). Because we aim to demonstrate the precision of synchronization detection, we add no time-variability to the model, the inference is  applied to a single block of data, and there is no spatial noise correlation with noise intensities $E_{11}=E_{22}=2$.
The dynamics of the phase difference is described as $\dot\psi=\Delta\omega-\eps \sin(\psi)+\xi_1(t)+ \xi_2(t)$, where $ \Delta \omega=\omega_2-\omega_1$ is the frequency mismatch and $\eps=\eps_{21}+\eps_{12}$ is the resultant coupling. It is evident that the analytic condition for synchronization, i.e.\ the existence of a stable equilibrium solution $\dot \psi<0$, is $\Delta \omega / \eps <1$. For $\eps_{12}=0.25$ (outside the Arnold tongue) the reconstructed map $M(\psi)$ (Fig.\ \ref{fig8:maps}(b)) after parameter inference has no root $M(\psi_e)=\psi_e$: hence the oscillators are not synchronized. When $\eps_{12}=0.35$, even though the system was inside the Arnold tongue, noise triggered occasional phase slips (see Fig.\ \ref{fig8:maps}(d)). We tested synchronization detection on the same signals using the methods already available in the literature, based on the statistics of the phase difference \cite{Tass:98,Mormann:00,Schelter:06}, but none of them was able to detect the presence of synchronization under these conditions.

For example, one of the most widely-used methods for synchronization detection \cite{Tass:98} gives a normalized index of 0.7539, well below the 0.9183 threshold (evaluated as the mean plus two SDs of surrogate realizations) for acceptance of synchronization.
In spite of the phase slips, our technique correctly detects the root $M(\psi_e)=\psi_e$ from the inferred parameters, revealing that the oscillators are {\it intrinsically} synchronized as shown in Fig.\ \ref{fig8:maps}(c): the phase slips are attributable purely to noise (whose inferred intensity is given by the matrix $E_{i,j}$), and not to deterministic interactions between the oscillators.

\subsubsection{Limit-cycle oscillators}\label{sec5_sub2_cph3:app_num}
To demonstrate the capabilities of the technique in tracing time-varying parameters, coupling functions, directionality and synchronization, we analyzed data from a numerical model of two coupled, non-autonomous, Poincar\'e oscillators subject to white noise,
\begin{equation}
\begin{split}
\dot x_i&= - r_i x_i  -\omega_i(t)\, y_i  + \varepsilon_{i}(t)\, q_i(x_i,x_j,t) +\xi_i(t),\\
\dot y_i&= - r_i y_i  +\omega_i(t)\, x_i  + \varepsilon_{i}(t)\, q_i(y_i,y_j,t) +\xi_i(t),\\
r_i&=(\sqrt{x_i^2+y_i^2}-1) \,\,\,\,\,\,\,\, i,j=1,2 \, .
\label{equ:num_model_app}
\end{split}
\end{equation}
We tested several possibilities for the parameters: while letting the frequencies $\omega_i$ and coupling parameters $\varepsilon_i$ be time-varying, we ran numerical experiments with the coupling function $q_i$ either fixed or time-varying.

\begin{figure}\begin{center}
\includegraphics[width=1\linewidth,angle=0]{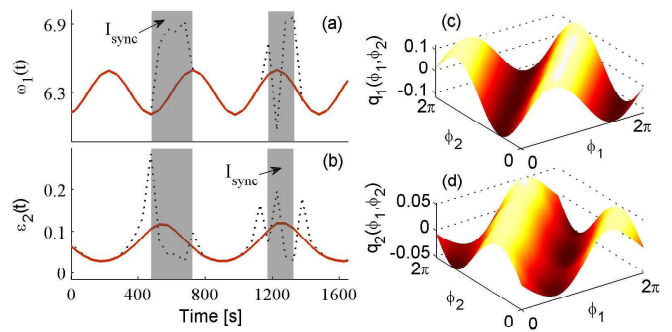}
\caption{(Color online) Extraction of time-varying parameters, synchronization and coupling functions from numerical data created by (\ref{equ:num_model_app}).  The frequency $\omega_1(t)$ (a) and coupling $\varepsilon_2(t)$ (b) are independently varied. The dotted and full lines plot the parameters when the two oscillators are synchronized for part of the time ($\varepsilon_{1}=0.3$), and not synchronized at all ($\varepsilon_{1}=0.1$), respectively. The regions of synchronization, found by calculation of the synchronization index, are indicated by the gray shaded regions. (c) and (d) show the coupling functions $q_1(\phi_1,\phi_2)$ and $q_2(\phi_1,\phi_2)$ for time windows centered at $t=350 s$. In both cases, the window length was $t_w=50 s$ and the coupling was $\varepsilon_{12}=0.1$.
}\label{fig9:num_sync}
\end{center} \end{figure}

As a first numerical experiment, we considered bidirectional coupling (1$\leftrightarrow$2), where the natural frequency of the first oscillator, and its coupling strength to the second one, vary periodically at the same time: $\omega_1(t)=\omega_1+\tilde A_1\sin({\tilde \omega_1}t)$ and $\varepsilon_{2}(t)=\varepsilon_{2}+\tilde A_2\sin({\tilde \omega_2}t)$. The other parameters were: $\varepsilon_{2}=0.1$, $\omega_1=2\pi\,1$, $\omega_2=2\pi\,1.14$, $\tilde A_1=0.2$, $\tilde A_2=0.13$, $\tilde \omega_1=2\pi\,0.002$, $\tilde \omega_2=2\pi\,0.0014$ and noise $E_{11}=E_{22}=0.1$. The coupling function was proportional to the difference in the state variables: $q_i(x_i,x_j,t)=x_i-x_j$ and $q_i(y_i,y_j,t)=y_i-y_j$ (the same coupling function was used for construction of Fig.\ \ref{fig1:tv_propg}, Fig.\ \ref{fig5:noises} and Fig.\ \ref{fig6:time_res}). The phases were estimated as the angle variable $\phi_i = \text{arctan}(y_i/x_i)$ (where $\text{arctan}$ is defined as the four-quadrant inverse tangent function).  With $\varepsilon_{1}=0.1$, in a state of no synchronization, the time-varying parameters $\omega_1(t)$ and $\varepsilon_{2}(t)$ are accurately traced as it can be seen in Fig.\ \ref{fig9:num_sync}(a) and (b).
The coupling amplitude of $\varepsilon_{1}=0.3$ corresponds to a state of intermittent synchronization, where the two oscillators are synchronized for part of the time. The precision of the reconstructed time-variable parameters is satisfactory during the non-synchronized intervals. During the synchronized intervals, however, the oscillators do not span sufficient phase-space to allow precise inference of the parameters (Fig.\ \ref{fig9:num_sync}(a) and (b), dashed lines). Within these synchronized intervals, the posterior probability distribution of the parameters was not peaked; however, it was sensibly different from zero only in that parameter region for which the corresponding noiseless dynamics is synchronized. Hence, despite the impossibility of accurate parameter tracking, the detection of a synchronized state ($s(c)=1$) is always precise (Fig.\ \ref{fig9:num_sync}(a) and (b), grey shaded regions).

The reconstructed sine-like functions $q_1(\phi_1,\phi_2)$ and $q_2(\phi_1,\phi_2)$ are shown in Figs.\ \ref{fig9:num_sync}(c) and (d) for the first and second oscillators, respectively. They describe the functional form of the interactions between the two Poincar\'e systems in Eq.~(\ref{equ:num_model_app}).  The reconstructed form of the coupling functions was evaluated dynamically for each block.

\begin{figure}[b]
\begin{center}
\includegraphics[width=1\linewidth,angle=0]{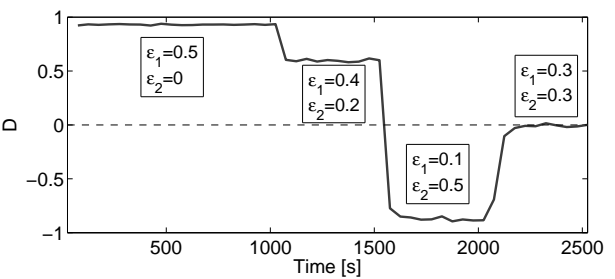}
\caption{Directionality of coupling $D$ for discretely time-varying coupling amplitudes $\varepsilon_1$ and $\varepsilon_2$. Different unidirectionally and bidirectionally coupled states are reached for different values of $\varepsilon_1$ and $\varepsilon_2$, as indicated by the square insets.} \label{fig10:num_dirc} \end{center}
\end{figure}

\begin{figure*} \begin{center}
\includegraphics[width=14cm,angle=0]{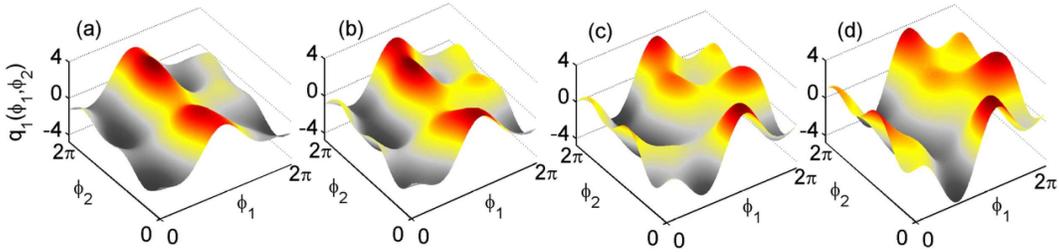}
\caption{(Color online) Time-evolution of the coupling function from model (\ref{equ:num_model_app}) with exponential time variations (\ref{equ:CF_tv}). (a)-(d) Coupling function $q_1(\phi_1,\phi_2)$ from the first oscillator for four consecutive time windows. The window length was $t_w=50 s$. For simplicity and clarity only the function $q_1(\phi_1,\phi_2)$ is shown. The behavior of $q_2(\phi_1,\phi_2)$ from the second oscillator was similar.}\label{fig11:num_CFs} \end{center}
\end{figure*}

Next, the method was applied to deduce the predominant direction of coupling as specified from the norm of the inferred coupling base parameters. To illustrate the precision of the directionality detection, the frequencies were now set constant, while both of the coupling strengths remained discretely time-varying. The parameters were $\omega_1=2\pi\,1.3$, $\omega_2=2\pi\,1.7$, $E_{11}=E_{22}=0.2$, and the coupling functions were, as in the previous example, $q_i(x_i,x_j,t)=x_i-x_j$ and $q_i(y_i,y_j,t)=y_i-y_j$. Synchronization was not reached, however, for these parameters. The couplings undergo changes at particular times, but otherwise remain constant, as shown in Fig.\ \ref{fig10:num_dirc}. The detected directionality index $D$ was consistent with the actual values.  Note that, for unidirectional coupling, $D$ does not quite reach unity on account of the noise.

To further investigate the ability to track subtle changes of time-varying coupling functions, we used the same model as in Eq.~(\ref{equ:num_model_app}) to generate a synthetic signal where the coupling functions are absolute values of the state difference to a power of the time-varying parameter:
\begin{equation}
\begin{split}
q_{x,i}(x_i,x_j,t)&=|(x_j-x_i)^{\nu(t)}|,\\
q_{y,i}(y_i,y_j,t)&=|(y_j-y_i)^{\nu(t)}|,
\label{equ:CF_tv}
\end{split}
\end{equation}
where $i,j=\{1,2\}$ and $i\neq j$. The exponent parameter varied linearly with time $\nu(t)=\{1\rightarrow3\}$, and the other parameters remained constant: $\omega_1=2\pi\,1$, $\omega_2=2\pi\,2.14$, $\varepsilon_{1}=0.2$, $\varepsilon_{2}=0.3$  and $E_{11}=E_{22}=0.05$. The reconstructed phase coupling functions $q_i(\phi_1,\phi_2)$ were calculated from the inferred parameters for the interacting terms of the base functions. The results for four consecutive windows are presented in Fig.\ \ref{fig11:num_CFs}. It can readily be seen that their complex form now is not constant, but varies with time. Comparing them in neighboring (consecutive) pairs: (a) and (b), then (b) and (c), then (c) and (d), one can follow the time-evolution of the functional form. Even though we can follow their time-variability, the two most distant functions Fig.\ \ref{fig11:num_CFs}(a) and (d) are of substantially different shapes. Note also that, beside their form, the functions' norm i.e.\ coupling strength also varies (cf.\ the height of the maxima in Fig.\ \ref{fig11:num_CFs}(a) and (d)).

Thus we have validated the technique on numerical models whose deterministic dynamics and time-variability were already known, thereby demonstrating the usefulness, precision and comprehensiveness of the method. We found that it can produce a good description of noise-induced phase-slips, synchronization, directionality and coupling functions even when the dynamics is subject to deterministic time-varying influences.

\subsection{Analogue simulations}\label{sec5_sub3_cph3:app_analg}
We also tested the technique on signals emanating from analog models. These are real, highly controllable, oscillatory systems and the noise on their signals is real rather than contrived, as in the case of numerical models. It is attributable to environmental disturbances, thermal fluctuations, and the inherent nonidealities of the circuit components. During the process of data acquisition and discretization, measurement noise can be introduced as well -- noise which has no links with the actual dynamics of the interacting oscillators. Such signals provide a good test of our analysis capabilities.
\begin{figure}[b!]
\begin{center}
\includegraphics[width=8cm,angle=0]{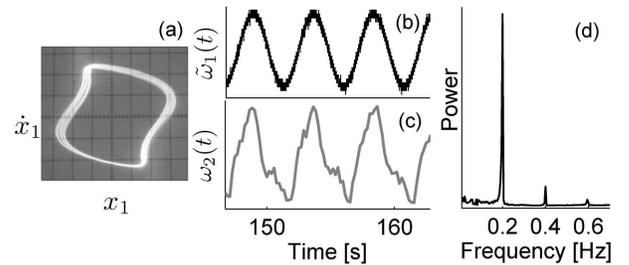}
\caption{Analysis of signals from an analogue simulation of the system (\ref{eq:VDPs}). (a) Phase portrait from the oscilloscope; (b) frequency $\tilde \omega_1(t)$ from the external signal generator; (c) detected frequency $\omega_2(t)$ of the second driven oscillator; (d) Fast Fourier Transform (FFT) of the detected frequency $\omega_2(t)$.}\label{fig12_analog} \end{center}
\end{figure}

We analyzed data from an analog experimental simulation of two coupled van der Pol oscillators. Details of the electronic implementation are given elsewhere \cite{Stankovski:12a}. The noise here arises mainly from the imperfections of the electronic components and there is also measurement noise.

Fig.\ \ref{fig12_analog}(a) shows the phase portrait derived from the first oscillator, with time-varying frequency, which drives the second oscillator
\begin{eqnarray}\label{eq:VDPs}
&\frac{1}{c^2}\ddot{x}_{1}-\mu_1(1-x_1^2)\frac{1}{c}\dot{x}_{1}+[
\omega_1+\tilde \omega_1(t)]^2x_1=0,\nonumber\\
&\frac{1}{c^2}\ddot{x}_{2}-\mu_2(1-x_2^2)\frac{1}{c}\dot{x}_{2}
+\omega_2^2x_2+\varepsilon(x_1-x_2)=0,\hspace{+0.8em}
\end{eqnarray}
where the periodic time-variability $\tilde \omega_1(t)=\tilde A_1 \sin({\tilde \omega}t)$ (Fig.\ \ref{fig12_analog}(b)) comes from an external signal generator. The parameters were $\varepsilon=0.7$, $\omega_1=2\pi\,15.9$, $\omega_2=2\pi\,17.5$, $\tilde A_1=0.03$, $\tilde \omega=2\pi\,0.2$ and $c=100$ is constant resulting from the analogue integration. The phases were estimated as $\phi_i=\arctan(\dot x_i/x_i)$.

For the given parameters the oscillators were synchronized, so that the second driven oscillator changed its frequency from being constant to being time-varying. Applying the inferential technique showed, correctly, that the oscillators were indeed synchronized ($s(c)=1$)  throughout the whole time period. The frequency of the driven oscillator was inferred as being time-varying Fig.\ \ref{fig12_analog}(c). Performing a simple FFT (Fig.\ \ref{fig12_analog}(d)) showed that $\omega_2(t)$ is periodic with period 0.2\,Hz (exactly as set on the signal generator).

Clearly, the technique reveals information about the nature and the dynamics of the time-variability of the parameters -- and is still able to do so using a more realistic signal than that from a numerical simulation.

\subsection{Cardiorespiratory interactions}\label{sec5_sub4_cph3:app_cvs}
Having tested our technique on two quite different kinds of synthetic data, we now apply it to a real physiological problem, to investigate the cardiorespiratory interaction. The analysis of physiological signals of this kind has already been found  useful in relation to several different diseases and physiological states (see e.g.\ \cite{Shiogai:10} and references therein). Transitions in cardiorespiratory synchronization have been studied in relation to an{\ae}sthesia \cite{Stefanovska:00a} and sleep cycles \cite{Bartsch:07}. It is also known that modulations and time-varying sources are present, and that these can affect the synchronization between biological oscillators \cite{Bracic:00c,Shiogai:10,Lewis:00}. For comprehensive and reliable analysis a technique is needed that is able, not only to identify the time-varying information, but which will allow evaluation of the interacting measures (e.g.\ synchronization and directionality), based solely on the information inferred from the signals. We will show that our technique meets these criteria.

We analyse cardiorespiratory measurements from human subject under an{\ae}sthesia. Their breathing rate was held constant, being determined by a respirator. For such systems the analytic model is unknown, in contrast to analogue and numerical examples, but the oscillatory nature of the signal is immediately evident. The instantaneous cardiac phase was estimated by synchrosqueezed wavelet decomposition \cite{Daubechies:11} of the ECG signal. Similarly, the respiratory phase was extracted from the respiration signal. The final phase time-series were reached after protophase-phase transformation \cite{Kralemann:07}. A more detailed explanation of the phase estimation procedure is given in Appendix \ref{app:phases}.

\begin{figure}[t!]
\begin{center}
\includegraphics[width=8cm,angle=0]{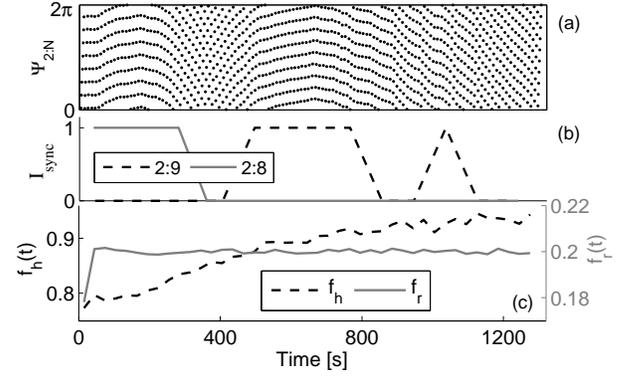}
\caption{Synchronization and time-varying parameters in the cardiorespiratory interaction. (a) Standard 2:$N$ synchrogram.  (b) Synchronization index for ratios 2:8 and 2:9 as indicated. (c) Time-evolution of the cardiac $f_h(t)$ and respiratory $f_r(t)$ frequency.  }\label{fig13:CVS_sync} \end{center} \end{figure}

Application of the inferential technique reconstructs the phase parameters that govern the interacting dynamics. Fig.\ \ref{fig13:CVS_sync}(c) shows the time-evolution of the cardiac and respiration frequencies. It is evident that the constant pacing of the breathing is well-inferred, and that the instantaneous cardiac frequency, i.e.\ ``heart rate variability'', increases with time. The inferred parameters, and their correlations, is used to detect the occurrence of cardiorespiratory synchronization and the corresponding synchronization ratio. The synchronization evaluation $I_{\rm sync}=s(c)\in\{0,1\}$, shown in Fig.\ \ref{fig13:CVS_sync}(b) reveals the occurrence of transitions between the synchronized and non-synchronized states, and transitions between different synchronization ratios: from 2:8 (i.e.\ 1:4) at the beginning to 2:9 in the later intervals. Because the evaluation of the synchronization state is based on all of the given details about the phase dynamics, the proposed method not only detects the occurrence of transitions, but also describes their inherent nature. The results for $I_{\rm sync}$ were consistent with the corresponding synchrogram shown in Fig.\ \ref{fig13:CVS_sync}(a), but provided a clearer and less ambiguous indication of synchronization.

\begin{figure}[t]
\begin{center}
\includegraphics[width=1\linewidth,angle=0]{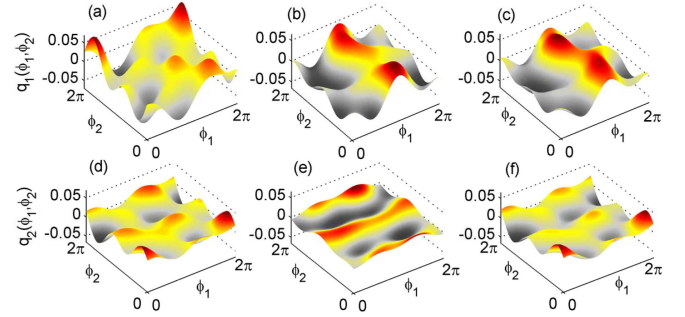}
\caption{(Color online) Coupling functions in the cardiorespiratory interaction calculated at different times. The cardiac and respiratory phases are represented by $\phi_1$ and $\phi_2$ respectively. (a)-(c) Coupling function $q_1(\phi_1,\phi_2)$ from the first oscillator, and (d)-(f) $q_2(\phi_1,\phi_2)$ from the second oscillator. The window time intervals were calculated at: $t=725$\,s for (a) and (d); $t=1200$\,s for (b) and (e); and at $t=1250$\,s for (c) and (f).
}\label{fig14_CVS_CFs} \end{center} \end{figure}

The functional relationships that describe the cardiorespiratory interactions are shown in Fig.\ \ref{fig14_CVS_CFs}. Evaluated for three different time windows, the upper figures (a)-(c) show the coupling function $q_1(\phi_1,\phi_2)$ from the cardiac oscillating activity, and the lower figures (d)-(f) show $q_2(\phi_1,\phi_2)$ from the respiration oscillator. The form of the functions is complex, and it changes qualitatively over time -- cf.\  Fig.\ \ref{fig14_CVS_CFs}(a) with (b) and (c), or (d) with (e) and (f). The influence from respiration to heart ($q_1(\phi_1,\phi_2)$) has a larger norm (i.e.\ coupling strength) than in the opposite direction, indicating that the predominant direction of coupling is from respiration to heart. One can also observe that $q_1(\phi_1,\phi_2)$ in (b) and (c) is of a fairly regular sinusoidal form with a strong influence from respiration. This arises from the contribution of those base functions describing the direct influence of respiration (for a detailed discussion see \cite{Iatsenko:12}). Furthermore, Fig.\ \ref{fig14_CVS_CFs} shows that the functional relationships for the interactions of an open (biological) system can in themselves be time-varying processes.

We conclude that the method is effective, not only when applied to digital and analogue synthetic signals, but also in the analysis of signals from human cardiorespiratory system. Unlike the synthetic signals, the cardiorespiratory signals are real, unpredictable, and subject to considerable time-variability. In this way, we were able to reconstruct the cardiac and respiratory frequency variabilities, estimate the direction of coupling, and detect the presence of cardiorespiratory synchronization and transitions between its different states. We also found that the form of the coupling functions themselves is a time-varying dynamical process.

\section{Generalization to networks of oscillators}\label{sc:networks}
\begin{figure*}[t!]
\includegraphics[width=0.8\linewidth,angle=0]{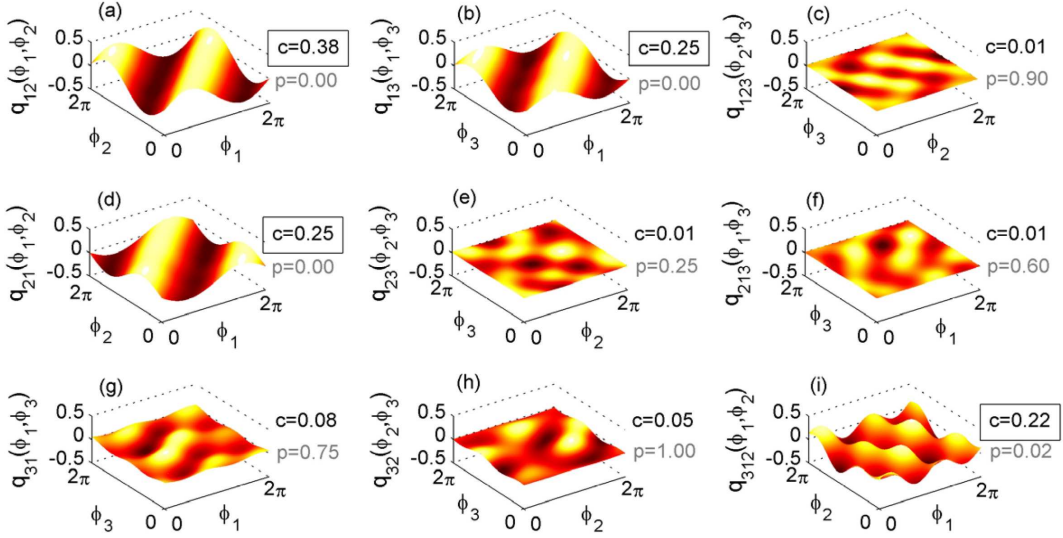}
\caption{\label{fig:net_CFs}(Color online) Coupling functions for
the interacting network (\ref{equ:num_model}). Each row represents
the influence on a specific oscillator: (a)-(c) on the first one,
(d)-(f) on the second, and (g)-(i) on the third. The notation is
such that e.g.\ $q_{23}$ represents the influence of the third
oscillator on the second, while $q_{213}$ represents the join
influences of the first and third oscillators on the second one.
The numbers on the right of each coupling function represent their
normalized coupling strength $c$ and the significance $p$-value.
The significant couplings are denoted with black squares. The
phases were estimated as $\phi_i=\arctan(y_i/x_i)$. The
parameter values are given in Table I.
 } \end{figure*}

Our parameter inference procedure can be applied with only minimal modification
to any number $N$ of interacting oscillators within a general coupled-network
structure.

The notation of Eq.\ (\ref{eq:phi}) is readily generalized for the $N$
oscillators, and the inference procedure, Eq.\ (\ref{eq:cD}), is then applied
to the corresponding $N$-dimensional phase observable.
For example, if one
wants to include all $k$-tuple interactions with $k\leq4$, then Eq.\ (\ref{eq:phi})
would be generalized into
\begin{equation}
\begin{split}
\dot \phi_i=& \omega_i + f_i(\phi_i) + \sum_j g^{(2)}_i(\phi_i,\phi_j) + \sum_{jk} g^{(3)}_{ijk}(\phi_i,\phi_j,\phi_k)\\ +& \sum_{jkl} g^{(4)}_{ijkl}(\phi_i,\phi_j,\phi_{k},\phi_{l}) + \xi_i.
\end{split}
\label{eq:phiN}
\end{equation}
Every function $g^{(k)}$ is periodic on the $k$-dimensional torus, and can be decomposed
in the sum of Fourier $k$-dimensional series of trigonometric functions.
Although, this decomposition is theoretically possible, it becomes less and less
feasible in practice as the number of oscillators and the number of $k$-tuples are
increased.  As a general approach, one could limit the number of base
functions to the most significant Fourier terms per $g^{(k)}$ functions;
but the task of finding the most significant component is algorithmically
demanding in itself. First, a very fast algorithm for the $k$-dimensional space
(such as \cite{Dolgov:12}) is required. Secondly,
since we have the value of each $\phi_i$ only at sparse values of the $\phi$'s that appear as argument in each
$g^{(k)}$, the algorithm should be adapted to deal with sparse, $k$-dimensional data,
as the one recently developed in \cite{Jiang:10}.
It is needless to say that, a part of the computational speed aspects,
the overall number of base functions should anyway be much less than the number of observed data.

In view of these difficulties, automatic selection of the most important Fourier terms to be used as base functions is hard to achieve on a network of more than just a few oscillators.
Known information about the system should be used to
reduce the number of base functions such that only those terms relevant to the
$N$-oscillator dynamics are included.

Other sub-procedures like the time-varying propagation, and the
noise inference, apply exactly as before. We note that the
computational power required increases very fast with
$N$, as discussed in Appendix \ref{app:comput}, which makes
the method unsuitable for the inference of large-scale networks.
However, for relatively small networks, a standard
high-performance personal computer will suffice for useful
inference.

We first demonstrate the inference on three interacting Poincar\'{e} oscillators subject to noise
\begin{equation}
\begin{split}
\dot x_i&= - r_i x_i  -\omega_i y_i  + \sum_j \varepsilon_{ij} x_j + \sum_{jk} \varepsilon_{ijk} x_j x_k  +\xi_i(t),\\
\dot y_i&= - r_i y_i  +\omega_i x_i  + \sum_j \varepsilon_{ij} y_j + \sum_{jk} \varepsilon_{ijk} y_j y_k +\xi_i(t),\\
r_i&=(\sqrt{x_i^2+y_i^2}-1) \,\,\,\,\,\,\,\, i,j,k=1,2,3 \, ,
\end{split}
\label{equ:num_model}
\end{equation}
where many of the coefficients $\varepsilon_{ij}$ and $\varepsilon_{ijk}$ are initially set to zero; but some are non-zero, such as when the first oscillator is pairwise coupled to the second and third oscillators. The second oscillator is coupled also to the first (forming a bidirectional interaction). The third oscillator is influenced by the join contribution from the first and second oscillators. The latter coupling means physically that part of the network (cluster) exhibits a common functional influence on the other oscillators. The inference of this cross-coupling is the direct benefit of  network (rather than pairwise) coupling detection.

\begin{table} \label{tab:par}
\begin{center}
\begin{tabular}{ c  c  c   c  c  c c c}
\hline
  Oscillator $i$  \vline & $f_i=\omega_i/2\pi$  \vline& Index $j$ & Index $k$ \vline& $\epsilon_{ij}$ & $\epsilon_{ik}$ & $\epsilon_{ijk}$ \vline& $E_i$\\
\hline \hline
1     & $1.1$& 2 & 3 & 0.3 & 0.2 & 0 & 0.1\\
2     & $0.27$ & 1 & 3 & 0.2 & 0 & 0 & 0.1 \\
3     & $3$ & 1  & 2 & 0 & 0 & 0.5 & 0.1\\
\hline
\end{tabular}
\end{center}
\caption{Parameters used for numerical simulation of systems (\ref{equ:num_model}). Note that the indexes $j$ and $k$ are introduced only for easier notation of the generic coupling amplitudes $\epsilon$.}
\end{table}

The inference of the three-dimensional phase variables from a
numerical simulation of the network (\ref{equ:num_model}) is
presented in Fig.\ \ref{fig:net_CFs}. The plots present the
specific forms of coupling function that govern the interactions
within the network. The coupling strengths are evaluated as
partial norms from the relevant base functions. Note that the
cross-couplings (c), (f) and (i) are shown for visual presentation
as functions dependent on two phases, whereas the coupling
strengths include also the base function dependent on all the
three phases. In order to determine whether the inferred couplings
are real or spurious, we conducted surrogate testing. The detected
couplings were tested for significance in respect of 100 couplings
evaluated from surrogate phases. Cyclic surrogates
\cite{Musizza:07,Vejmelka:08} were generated from each of the
phases, randomizing the temporal cross correlations, while
preserving the frequencies and statistical characteristics
unchanged.

Recently, Kralemann et al.\ \cite{kralemann:11} discussed the notion of
effective and structural connectivity in networks. Effective couplings are
those that are detected, while not present in the original structure  e.g.\
indirectly-induced coupling. In our numerical examples, the structural
couplings are the parameters from the numerical simulation, while the effective
are those evaluated as partial norms from the inferred parameters. The
question posed was: are the effective couplings real, or are they artifacts?
Our analysis showed that when one applies appropriate surrogate testing, the
technique is able to distinguish the structural couplings as being significant.
The resultant coupling strengths and significance $p$-values in Fig.\
\ref{fig:net_CFs} suggest that the connectivity (black-boxed couplings) of the
network (\ref{equ:num_model}) was inferred correctly. Note that some relations
as in (g) have relatively large strength, even though they are less significant
then some lower couplings as in (e). If the possibility of effective couplings
cannot be excluded, then our technique (with use of surrogate testing) provides
a consistent way of inferring the true structure of the network. It is
also important to note that the coupling strength is evaluated as a partial
norm and its value is not necessarily equal to the
structural value, but is only proportional to it. When one infers complex networks, it is not only important
what the structural coupling value is, but also how the oscillators are coupled
and what are the coupling functions between the oscillators.

More importantly, the use of our method allows one to follow the
time-variability of the structural and functional connectivity
within the network. This is especially important when inferring
the interactions of biological oscillators, for which it is known
that the dynamics is time-varying
\cite{Stankovski:12b,Stefanovska:99a,Rudrauf:06}. To illustrate
the latter we infer the following network of four phase
oscillators subject to white Gaussian noise
\begin{equation}
\begin{split}
\dot \phi_1&= \omega_1+a \sin(\phi_1)+\varepsilon_{13}(t) \sin(\phi_3)+\varepsilon_{14}(t) \sin(\phi_4)+\xi_1(t)\\
\dot \phi_2&= \omega_2+a \sin(\phi_2)+\varepsilon_{21}(t) \sin(\phi_2-\phi_1)+\xi_2(t)\\
\dot \phi_3&= \omega_3+a \sin(\phi_3)+\varepsilon_{324}(t) \sin(\phi_2-\phi_4)+\xi_3(t)\\
\dot \phi_4&= \omega_4+a \sin(\phi_4)+\varepsilon_{42}(t) \sin(\phi_2)+\xi_4(t) \, .
\label{equ:num_model2}
\end{split}
\end{equation}
Note that, because the coupling strengths are functions of time,
we were effectively changing the structural connectivity of the
network by varying their values. The parameter values for the
simulations were: $\omega_1=2\pi\,1.11$, $\omega_2=2\pi\,2.13$,
$\omega_3=2\pi\,2.97$, $\omega_1=2\pi\,0.8$, $a=0.2$, and noise
strengths $E_i=0.1$. The couplings were varied discreetly in three
time-segments, as follows. (i) For 0-500s: $\varepsilon_{13}=0.4$,
$\varepsilon_{14}=0.0$, $\varepsilon_{324}=0.4$ and
$\varepsilon_{42}=0.4$. (ii) For 500-1000s: $\varepsilon_{13}=0$,
$\varepsilon_{14}=0.35$, $\varepsilon_{324}=0$ and
$\varepsilon_{42}=0.4$. (iii) For 1000-1500:
$\varepsilon_{13}=0.45$, $\varepsilon_{14}=0.35$,
$\varepsilon_{324}=0$ and $\varepsilon_{42}=0$. The coupling
$\varepsilon_{21}$ was continuously varied between
$0.5\rightarrow0.3$. Note also that in Eq.
(\ref{equ:num_model2}) the coupling functions are qualitatively
different i.e.\ the arguments in the sine functions are not the
same for each oscillator. For example the coupling functions for
$\varepsilon_{13}$, $\varepsilon_{14}$ and $\varepsilon_{42}$ have
one phase argument, while the coupling functions for
$\varepsilon_{21}$ and $\varepsilon_{324}$ have the phase
difference as their argument. The last two are additionally
different because the coupling function with $\varepsilon_{21}$
for the second oscillator contains its own phase $\phi_2$ in the
phase difference.

\begin{figure}
\includegraphics[width=0.95\linewidth,angle=0]{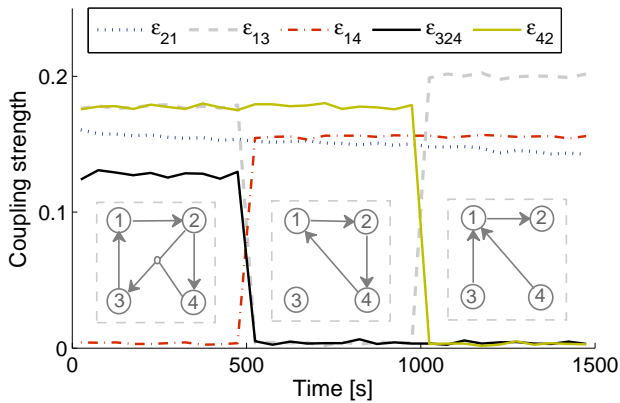}
\caption{\label{fig:net_TV} (Color online) Inference of time-varying coupling structure for the network (\ref{equ:num_model2}). The color/grayscale code for the couplings is presented in the box at the top, where $\varepsilon_{21}$ is represented by a dotted line, $\varepsilon_{13}$ by a dashed line, $\varepsilon_{14}$ by a dash-dotted line, $\varepsilon_{324}$ by a bold full line and $\varepsilon_{42}$ by a light full line. The four couplings $\varepsilon_{13}$, $\varepsilon_{14}$, $\varepsilon_{324}$ and $\varepsilon_{42}$ were held constant at different values within three time segments each of length 500\,s. However, $\varepsilon_{21}$ was varied continuously through the whole time interval. For each segment the structure of the network is presented schematically on the diagrams in the dashed grey boxes. The parameters are given in the text.}
\end{figure}

 The results are presented in Fig.\ \ref{fig:net_TV}. In the
first interval (0-500s) we inferred three pairwise coupling
amplitudes $\varepsilon_{21}$, $\varepsilon_{13}$ and
$\varepsilon_{42}$, and also one joint coupling
$\varepsilon_{324}$ which results from the joint influences of the
second and fourth oscillators on the third one. The schematic
diagram above the 0-500s time interval represents the structural
connectivity, where the arrows indicate the direction of influence
between the oscillators. On the transition to the second interval
(500-1000s) two of the couplings $\varepsilon_{13}$,
$\varepsilon_{324}$ disappear and one new one $\varepsilon_{14}$
appears. This change occurs discretely at the instant of
transition between the two regions. Two couplings continue to
exist: $\varepsilon_{42}$ at a constant level, while
$\varepsilon_{21}$ decreases linearly and continuously. The second
schematic diagram shows the structure of the network in this
interval. Comparing the diagrams describing the first two
intervals one may note that the method infers correctly the
couplings and their time-variability, and by doing so it infers
the network connectivity even though it is changing with time.
Similarly the transition to the third interval (1000-1500s)
detects the alternations of two couplings $\varepsilon_{42}$ and
$\varepsilon_{13}$. This leads to a new connectivity state of the
network, as presented in the third schematic diagram. The results
from the whole time span demonstrate that the method follows the
time-variability of the couplings effectively and precisely. The
dynamical variations are taking the network structure through
various different connectivity states, and the different
topologies are detected reliably throughout their time-evolution.

\section{Conclusions} \label{sc:conclusion}
Starting from the perspective of dynamical systems inference, we have built an algorithm able to detect synchronization, to describe the functional form of the mutual interactions between oscillators, and to perform such tasks successfully in the presence of a time-evolving dynamics.

The algorithm differs substantially from earlier approaches with respect both to synchronization detection capabilities, and to the estimation of coupling and directionality. Most other techniques are based on information flow (e.g.\ transfer entropy, or Granger causality) providing them with great generality. While limiting ourself to the hypothesis of continuous time differential equations driving the dynamics (correspondingly restricting the domain of applicability), we can optimally exploit the benefits of this assumption. Unlike all other approaches, our technique does not require the observable to fill the domain of the probability density function at equilibrium. Thus, in oscillators with a limit cycle (Van der Pol, Fitz-Hugh Nagumo, Poicar\`e, etc...) even one single extreme path is sufficient to characterize the parameters of the dynamics. Hence, we can determine uniquely the limit-time equilibrium distribution, i.e.\ the Fokker-Plank equation associated with the SDE. Thus an immediate advantage is that we can extract the same information from a fraction of the volume of data that is typically required by earlier methods. Because a very wide range of natural and artificial systems are describable in terms of continuous time differential equations (e.g.\ oscillatory processes in nature, mechanical systems, analogue voltage systems), the loss of generality in our approach is actually minimal, compared to the advantage gained in terms of informational efficiency.

We have applied the algorithm successfully to a representative classes of oscillators, testing it on synthetically-generated data created from various models, and on data from an analogue circuit device with known dynamics. In each case, we were able to demonstrate the precision of parameter detection, the temporal precision of synchronization detection, and the accuracy of directionality identification.

We have also demonstrated the efficacy of the technique in relation to cardiorespiratory time series data. Synchronization phenomena were already well-known in such systems, but the details of functional coupling were not.  From the inferred parameters we were able to reconstruct the extent of the cardiac and respiratory variability, estimate the direction of coupling, and detect the presence of and type of intermittent cardiorespiratory synchronization.

Because the whole enterprize is built on an inference algorithm for an $N$-dimensional dynamical system, the technique was readily extensible to the study of a network of oscillators whose parameters and coupling functions may be changing in time. An example of such application we considered a network of Poincar\'e oscillators, generated by numerical simulation. We were able to demonstrate  effective coupling detection, cross-validating the results by surrogate testing.

Although the implementation itself might see future improvements (e.g.\ in terms of speed of calculation, or automatic base function selection), it is worth emphasizing that the method allows one to designate which  components of the system are expected to be time-variable. Such selection is optional, but it provides an effective means by which to incorporate previous knowledge available for any particular system, and enables the algorithm to adapt itself optimally to the externally-imposed constraints.

Given the advantages that the dynamical approach offers in tackling synchronization detection and coupling identification, we believe that the framework presented above will be found valuable for a wide range of future applications.

\appendix

\section{Fixed point algorithmic check} \label{app:fixedpoint}
The procedure of synchronization detection between two oscillators generating phase time-series
reduces to the investigation of synchronization of the synthetic phase model using the parameters returned by the Bayesian algorithm. To calculate $s(c)$ for any of the sampled parameter sets,  one can proceed as follows:

\begin{itemize}

\item[(i)] From an arbitrary fixed $\zeta$, and for an arbitrary $\psi_0$, integrate numerically (using the standard fourth-order Runge-Kutta algorithm) the dynamical system prescribed by the phase base function (Eq.\ (\ref{eq:phiF}) without the noise) for one cycle of the toroidal coordinate, obtaining the mapped point $M(\psi_0)$.

\item[(ii)] Repeat the same integration for multiple $\psi_i$ coordinates next to the initial one, obtaining the map $M(\psi_i)$

\item[(iii)] Based on finite difference evaluation of $dM/d\psi$, use a modified version of Newton's root-finding method to analyse the function $M(\psi)-\psi$. The method is modified by calculating $M$ at the next point $\psi_{n+1}$ such that $$\psi_{n+1}= \psi_n + 0.8\times | (M(\psi_n)-\psi_n)/(M'(\psi_n)-1))|.$$ The coefficient $0.8$ is an arbitrary constant that we found to be particularly efficient for solution of the problem. Note that in this version, Newton's method can only test the function by moving forward; in actual fact (a) the existence of the root is not guaranteed; and (b) we are not interested in the root itself but only in its existence.

\item[iv)] If there is a root, $s(c)=1$ is returned. If a root is not found, $s(c)=0$ is returned.

\end{itemize}

\section{Reliable phase estimation from ECG and respiration signals} \label{app:phases}
In order to infer the phase dynamics, one needs to have good estimates of the phases from the observable time-series. This is even more important when the oscillatory dynamics is time-varying and the analysis requires instantaneous phases. Potential difficulties for phase estimation arise when the signals emanate from complex, highly nonlinear and/or mixed-mode oscillatory dynamics. Although the phase from the respiration signal is relatively easy to detect, obtaining the instantaneous phase from the ECG signal is considerably more difficult.

We used the synchrosqueezed wavelet transform \cite{Daubechies:11} to estimate phases from the complex and nonlinear ECG and respiration signals. Given a signal $g(t)$ we first calculate its wavelet transform in the scale-time domain $(s,t)$,
\begin{equation}\label{equ:wt2}
W(s,t)=\int_{-\infty}^{\infty} \bar{\Psi}_{s,t}(u) \cdot g(u)du,  
\end{equation}
where the $\bar{\Psi}$ represents the complex conjugate of the mother wavelet $\Psi$
\begin{equation}\label{equ:wt1}
\Psi_{s,t} (u)= |s|^{-1/2} \cdot \nu \left(\frac{u-t}{s}\right). 
\end{equation}
We use the Morlet mother wavelet
\begin{equation}
\nu (u)= \frac{1}{\sqrt[4]{\pi}}\text{ } e^{-i 2\pi f_0 u}\cdot e^{-u^2/2}, \nonumber
\label{equ:wavelet}
\end{equation}
where the central frequency was set to be $f_0=1{\rm \, Hz}$.

The synchrosqueezed transform aims to ``squeeze'' the wavelet around the intrinsic frequency in order to provide better frequency localization. For any $(s,t)$ for which $W(s,t)\neq0$, a candidate instantaneous frequency for the signal $g$ can be calculated as
\begin{equation}\label{equ:freq}
\omega_g(s,t)=-i\frac{\frac{\partial}{\partial t}W_g(s,t)}{W_g(s,t)}.
\end{equation}
The information from the time-scale plane is transferred to the time-frequency plane, according to a map $(s,t)\rightarrow(\omega_g(s, t),t)$, in an operation called synchrosqueezing. The synchrosqueezed wavelet transform is then expressed as
\begin{equation}\label{equ:sqz}
T_g(w,t)=\int_{A(t)}W_g(s,t)s^{-3/2}\delta(\omega(s,t)-\omega)ds,
\end{equation}
where $A(t)=\{ a;W_g(s,t)\neq 0\}$, and $\omega(s, t)$ is as defined in (\ref{equ:freq}) above, for $(s, t)$ such that $s\in A(t)$. The complex (as with real and imaginary values) nature of the synchrosqueezed transform allows one to extract the phase of the signal as the angle of the transform
\begin{equation}\label{equ:sq_phase}
\theta(t)=\angle\big[ \sum_{k} T_g(\omega,t)(\Delta \omega) \big ].
\end{equation}
The transform's great advantage lies in its ability to determine instantaneous characteristics from complex signals with non-harmonic waveforms.

Evaluated through such a procedure the phases $\theta(t)$ may, however, be observable-dependent
and non-universal, i.e.\ they can retain premises resulting from the phase-detection technique (in this case the synchrosqueezed transform) but not from the genuine phases. They are therefore treated as protophases, and a special technique is applied to transform the protophases into true phases $\phi(t)$ that are independent of the observable and are universally defined \cite{Kralemann:07}. The transformation can be written as
\begin{equation}\label{equ:proto}
\phi=\theta + \sum_{n\neq0} \frac{S_n}{in}(e^{in\theta}-1),
\end{equation}
where $S_n$ are coefficients from a Fourier expansion of the averaged phase relationships. For further detail see \cite{Kralemann:07}.

\section{Computational speed consideration} \label{app:comput}
For a sufficiently large number of parameters $M$, the complexity of the
algorithm for parameter estimation is substantially dominated by O$(M^3)$,
which is of the order of the time required for an $M$-sized matrix inversion.
For a network of $N$ oscillators, if one consider all possible pairwise connections
then $M\propto c_1 N^2$, where $c_1$ is a proportionality coefficient to account for
by the truncation order of the fourier decomposition.
Similarly, if one considers all pairwise connections and every double
connection up to a truncation order of $c_2$, then $M\propto c_1 (N^2 + c_2^{2} N \binom{N}{2})$.
With recursive reasoning, if one considers all the $k$-tuples with $k$ up to $P$,
each with a truncation order of $c_k$,
then the number of coefficients would grow as
$
M\propto c_1 N \times \sum_{k=1}^{P} c_k^k { \binom{N}{k} }  \,\,.
$
It is clear that even for a modest network, considering just a few $k$-tuples of
possible connections would be unfeasible in practice.
Very careful selection of the base functions is therefore always to be recommended.

\end{document}